\documentclass[sigconf]{acmart}

\usepackage{physics}
\usepackage{subcaption}
\usepackage{graphicx}
\usepackage{fixltx2e}
\usepackage{amsmath}
\usepackage{textcomp}
\usepackage{enumitem} 
\usepackage{gensymb}
\usepackage{algorithm}
\usepackage{caption}
\allowdisplaybreaks
\usepackage[noend]{algpseudocode}
\usepackage{tikz}

\newcommand*\circled[1]{\tikz[baseline=(char.base)]{
            \node[shape=circle,draw,inner sep=0.5pt] (char) {#1};}}
\AtBeginDocument{%
  }

\copyrightyear{2024}
\acmYear{2024}
\setcopyright{acmlicensed}\acmConference[DAC '24]{61st ACM/IEEE Design Automation Conference}{June 23--27, 2024}{San Francisco, CA, USA}
\acmBooktitle{61st ACM/IEEE Design Automation Conference (DAC '24), June 23--27, 2024, San Francisco, CA, USA}
\acmDOI{10.1145/3649329.3655923}
\acmISBN{979-8-4007-0601-1/24/06}

\acmConference[DAC'24]{June 23--27,
  2024}{San Francisco, CA, USA}

\begin{document}

\title{Fast Virtual Gate Extraction For Silicon Quantum Dot Devices}

\author{Shize Che, Seongwoo Oh, Haoyun Qin, Yuhao Liu, Anthony Sigillito, Gushu Li}
\email{{shizeche, ohseong, qhy, liuyuhao, asigilli, gushuli}@seas.upenn.edu}
\affiliation{
  \institution{University of Pennsylvania, USA}
    \country{}
}

\renewcommand{\shortauthors}{Che et al.}

\begin{abstract}
  Silicon quantum dot devices stand as promising candidates for large-scale quantum computing due to their extended coherence times, compact size, and recent experimental demonstrations of sizable qubit arrays. Despite the great potential, controlling these arrays remains a significant challenge. This paper introduces a new virtual gate extraction method to quickly establish orthogonal control on the potentials for individual quantum dots. Leveraging insights from the device physics, the proposed approach significantly reduces the experimental overhead by focusing on crucial regions around charge state transition. Furthermore, by employing an efficient voltage sweeping method, we can efficiently pinpoint these charge state transition lines and filter out erroneous points. Experimental evaluation using real quantum dot chip datasets demonstrates a substantial 5.84× to 19.34× speedup over conventional methods, thereby showcasing promising prospects for accelerating the scaling of silicon spin qubit devices.
\end{abstract}

\ccsdesc[500]{Computer systems organization~Quantum computing}
\ccsdesc[300]{Hardware~Electronic design automation}

\keywords{Quantum Dot, Device Tuning, Virtual Gate}


\maketitle

\section{Introduction}


Electron spin qubits in gate-defined silicon quantum dots are one of the most promising candidates to implement large-scale quantum computers due to long coherence times ~\cite{yoneda_quantum-dot_2018, tyryshkin_electron_2012}, small size~\cite{siliconSpinReview}, and fast single- and two-qubit operations~\cite{Xue, Noiri, Mills}. In addition, silicon quantum dots can be fabricated \textit{en masse} using the existing semiconductor fabrication infrastructure, which offers enormous potential for rapid scaling~\cite{siliconIndustry}. 
As of now, most of the DiVincenzo's criteria~\cite{divincenzo2000physical} for a quantum computer have been satisfied for silicon spin qubits: long coherence times ~\cite{yoneda_quantum-dot_2018, tyryshkin}, 
state preparation~\cite{Mills, watson_programmable_2018, huang}, 
high-fidelity universal gates ~\cite{zajac_resonantly_2018, watson_programmable_2018, Mills, huang_fidelity_2019, philips_universal_2022}, 
and high-fidelity readout ~\cite{Mills4, PhysRevX.8.021046, takeda_rapid_2023}. 
Recently, the quantum dot chips of 12 qubits and 16 qubits have been experimentally demonstrated by Intel~\cite{TunnelFall} and Delft~\cite{Borsoi}, respectively.

While fabricating large-scale quantum dot arrays is relatively straightforward, controlling such large-scale arrays is a major hurdle.
Quantum dot device tuning is a multi-stage process in which the voltages applied to metallic electrodes are varied until the device supports individual electrons beneath specific gates. Depending on the qubit encoding, the ideal electronic charge configuration can vary. This process becomes increasingly complicated and time-consuming as the number of qubits increases.
Currently, it can take hours of human effort to achieve a satisfying voltage configuration.
This has motivated the development of computer-aided tuning techniques targeting various stages of the tuning process~\cite{ziegler2023tuning, Mills2, moon2020machine, Ziegler, Van_diepen, Giovanni}.



This paper focuses on a critical step in the tuning process, establishing independent control over the energy potential of each dot in a device. Ideally, a gate voltage only affects the potential level of its corresponding dot, but due to cross-capacitance, it also affects the potential levels of nearby dots. A widely adopted technique for establishing such ``one-to-one" control is constructing the so-called virtual gate with a linear combination of the capacitively coupled gates to a given dot ~\cite{baart_single-spin_2016}. This virtual gate technique has been successfully applied to quantum dot array devices 
\cite{Volk, Mills3}.


Existing automation techniques for virtual gate extraction rely on analyzing complete \textit{charge stability diagrams}, deploying computer vision techniques such as the Hough transform~\cite{Mills2, Giovanni}, or more advanced techniques such as convolutional neural network~\cite{Ziegler}. However, the techniques' reliance on low-noise charge stability diagrams dramatically limits their scalability since obtaining a charge stability diagram on an actual device often takes minutes. The number of diagrams scales linearly with the number of dots, creating multi-hour overhead on devices with merely tens of qubits.
As the number of qubits grows, a fast automated procedure
for virtual gate extraction will be vital for scaling up silicon spin qubit devices.



In this paper, we make a key observation that there is a large redundancy in existing virtual gate extraction methods. 
Most of these data points in the charge stability diagram are not necessary, and only those data points on or close to the charge state transition lines contribute to constructing the virtualization matrix.
Suppose we can locate the region of transition lines. In that case, we can probe only the points near them to significantly reduce the required experimental data and thus accelerate the virtual gate extraction. 


To this end, this paper proposes a fast virtual gate extraction method. By making reasonable assumptions based on the device physics, our method can locate the transition lines and construct the virtualization matrices with much fewer experimental requirements.
\textbf{First,} the device physics puts constraints on the possible values of the transition line slopes. Thus, the transition lines can only appear on a specific region in the entire charge stability diagram.
By quickly locating this specific region, we can greatly reduce the number of voltage configurations we need to experimentally test.
\textbf{Second,} we further propose an efficient voltage configuration sweeping method to adaptively narrow our data probing to the critical area around the transition lines to further reduce the experimental requirements.

We evaluate the proposed fast virtual gate method on charge state datasets of real quantum dot chips from the qflow dataset~\cite{qflow}.
Experimental results show that our method can outperform the baseline method using Hough transform with about $5.84\times$ to $19.34\times$ speedup with an even higher success rate. 
The speedup mostly comes from the reduction in the
number of data points probed as our method only requires about $10\%$ of the entire CSD data to locate the transition lines.

\section{Background}

In this section, we introduce the basics of quantum dot technology and gate voltage tuning. 
Note that the term `gate' in the rest of this paper refers to the hardware gate electrode fabricated on the chip rather than the unitary transformations in the quantum algorithm/software.

\begin{figure}[t]
  \centering
  \begin{subfigure}{0.49\columnwidth}
    \centering
    \includegraphics[width=0.93\linewidth]{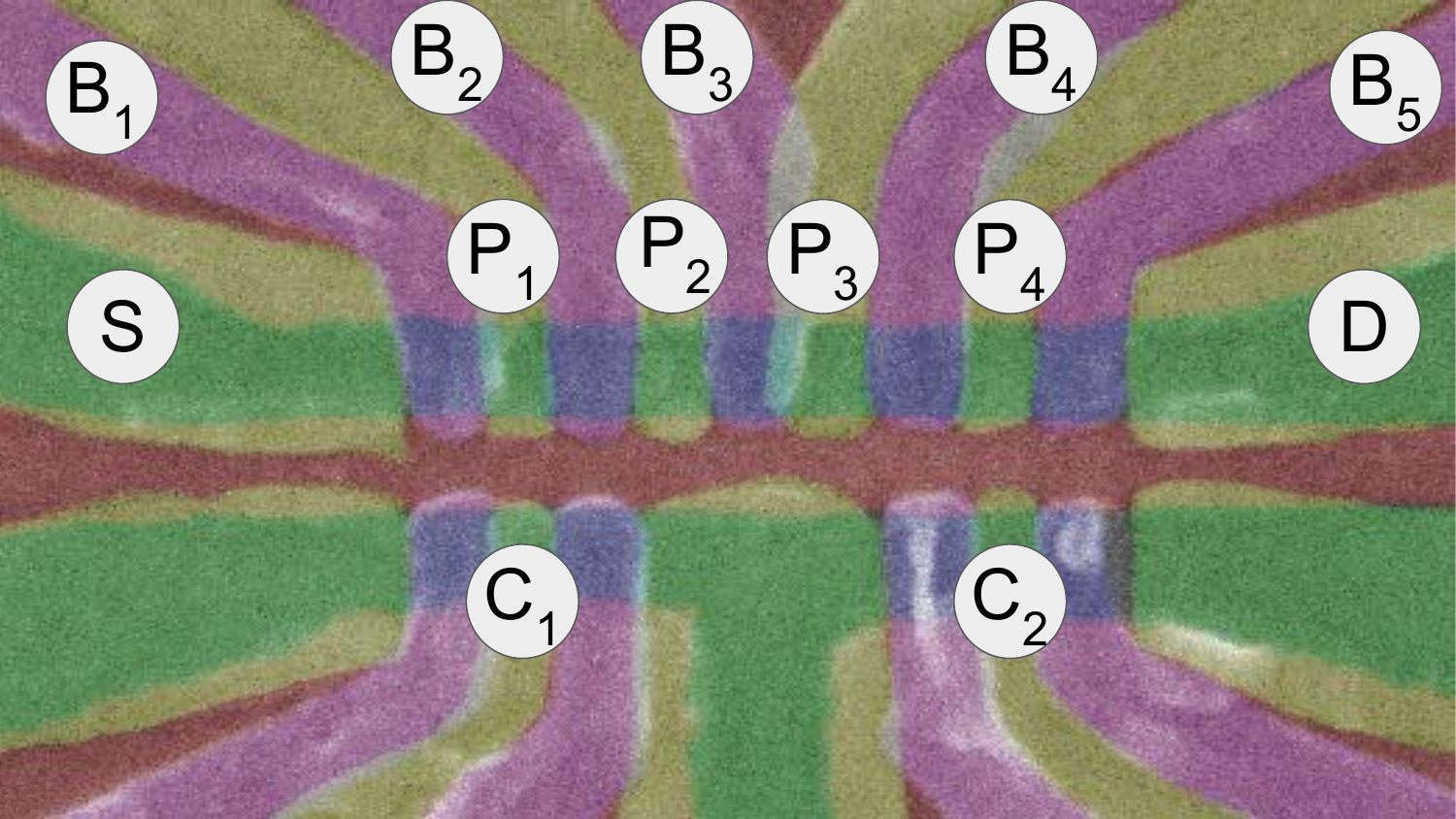}
  \vspace{-5pt}
  \caption{}
  \end{subfigure}
  \hfill
  \begin{subfigure}{0.49\columnwidth}
    \centering
    \includegraphics[width=0.93\linewidth]{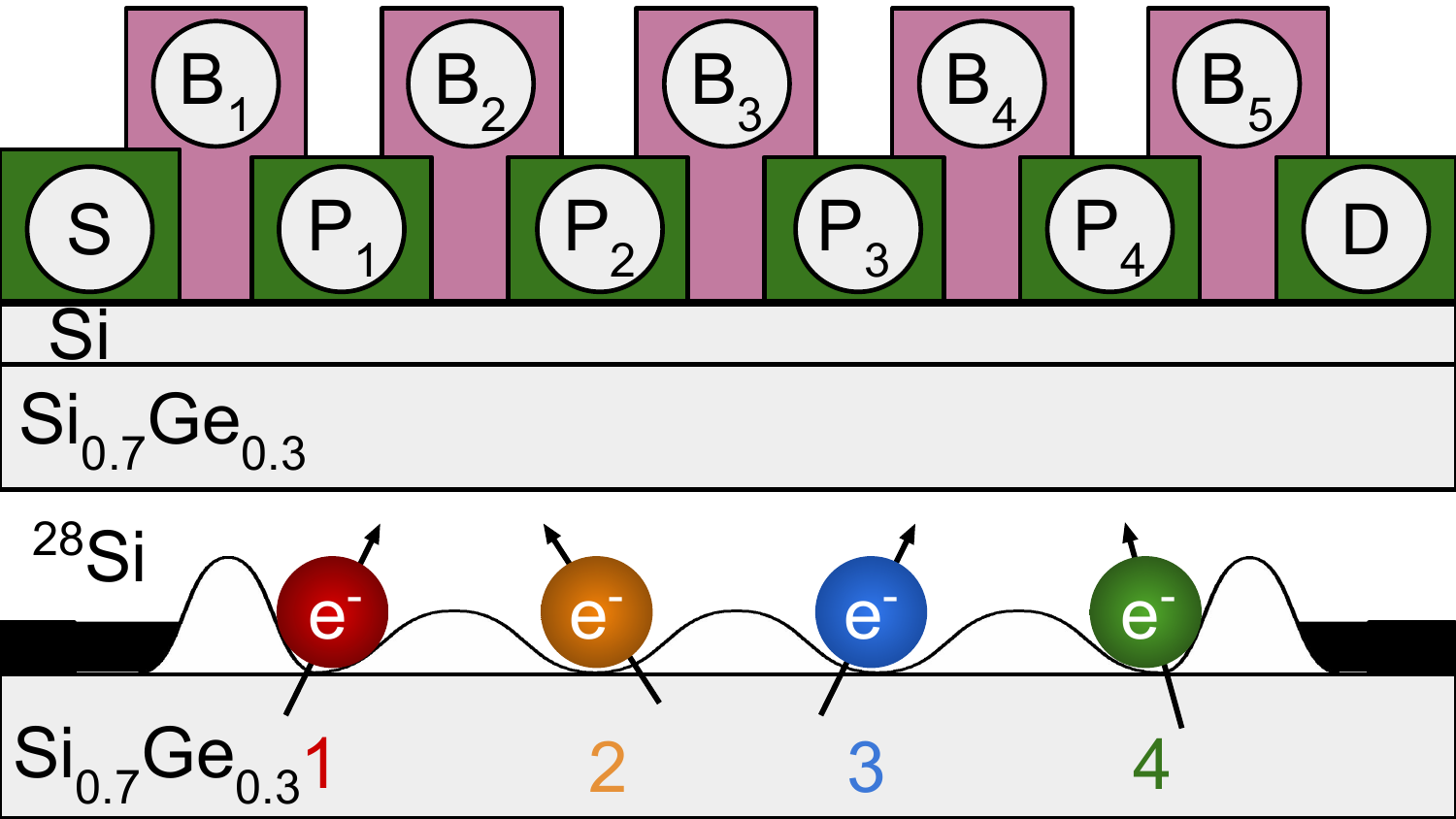}
  \vspace{-5pt}
  \caption{}
  \end{subfigure}
  \vspace{-10pt}
  \caption{(a) Top-view scanning electron microscopy of a qua-druple-dot device, (b) Lateral-view of the device's dot-side}
  \vspace{-15pt}
  \label{fig:quantum-dot}
\end{figure}

\subsection{Quantum Dot}
The quantum dot devices under consideration in this paper have the structure shown in Figure~\ref{fig:quantum-dot} (a). 
On one side of this device (the upper half of Figure~\ref{fig:quantum-dot} (a)), it consists of a source (marked by $S$), a drain (marked by $D$), a set of plunger gates (marked by $P_1\sim P_4$) and a set of barrier gates (marked by $B_1\sim B_5$). By adjusting the gate voltages, 
one can trap electrons under the gates,
forming the so-called ``dots''. On the opposite side (the lower half of Figure~\ref{fig:quantum-dot} (a)) of the device, there are two charge sensors (C\textsubscript{1} and C\textsubscript{2}), both of which are single quantum dots. The charge sensors' conductance is sensitive to changes in the local electrostatic potential. It can, therefore, detect changes in the number of electrons in the dots through current change in the sensors. 

The upper half of Figure~\ref{fig:quantum-dot} (b) 
shows the cross-sectional view of the plunger ($P_1\sim P_4$) and barrier gates ($B_1\sim B_5$) 
as well as the additional layers and the trapped electrons below the gates 
(charge sensors are on the opposite side and not shown). 
The device has multiple layers made of different semiconductor materials. The electrons are sandwiched in the middle layer. They are subject to the electrostatic potential (represented by the curvy line in the lower half of Figure~\ref{fig:quantum-dot} (b)) generated from tuning the gate voltages. A desirable potential profile, like the one in Figure~\ref{fig:quantum-dot} (b), creates four dots labeled from 1 to 4 under each plunger gate. In Figure~\ref{fig:quantum-dot} (b), each dot confines and stabilizes one electron, forming the Loss-DiVincenzo qubits ~\cite{DiVincenzo}, where $\ket{0}$ and $\ket{1}$ are encoded as the spin up and down of the electron, respectively.


\subsection{Charge Stability Diagram}
\label{sec:virtual-gate}

\begin{figure}[h]
  \centering
  \includegraphics[width=\columnwidth]{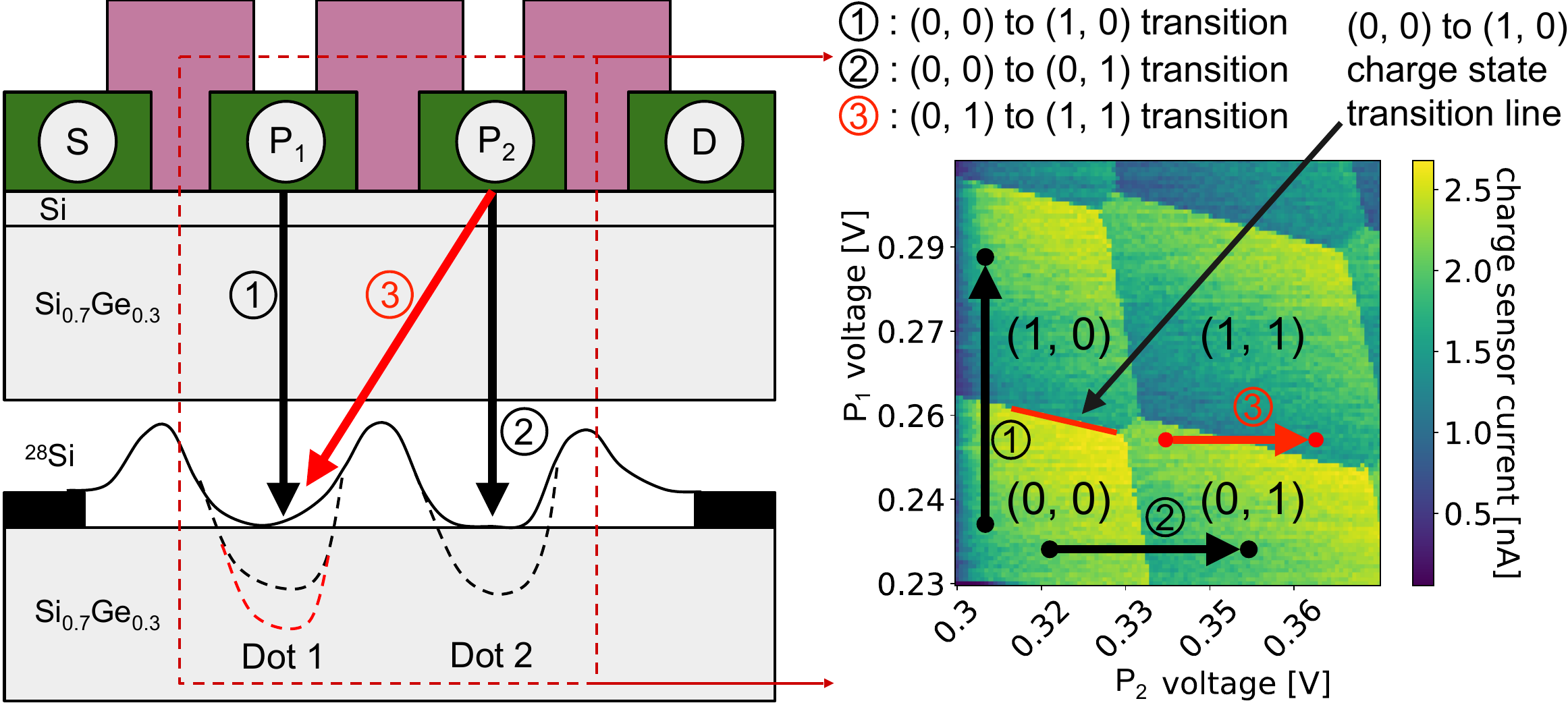}
   \vspace{-10pt}
  \caption{Example charge stability diagram of a double quantum dot device}
  \vspace{-10pt}
  \label{fig:dqd-charge-stability}
\end{figure}

To understand how gate voltage tuning can trap the electrons, an example of a double-dot device is shown in Figure~\ref{fig:dqd-charge-stability}.
On the left is the cross-sectional view of this device and the electrostatic potential designed to trap individual electrons.
This device has two plunger gates, and electrons are trapped under each plunger gate.
Since an electron carries a negative charge, increasing $P_1$ or $P_2$ voltage deepens the potential level at the dot 1 or 2 (from the black solid line to the black dashed line), respectively.
Then, we can increase/decrease the number of electrons in the dot by adjusting its corresponding plunger gate voltage to raise/lower the energy potential.
However, adjusting the plunger gate voltage also has an unwanted effect, the cross-capacitance,
which will affect the potential at the nearby dots by a non-negligible amount. 
For example, the increased $P_2$ voltage also pushes the potential level of dot 1 further to the red dashed line, which is deeper than the desired black dashed level.

\textbf{Charge Stability Diagram} The effect of nearby plunger gates on the charge state can be visualized in a Charge Stability Diagram (CSD).
The CSD of the example double-dot device is on the right of Figure~\ref{fig:dqd-charge-stability}. 
A CSD represents the charge sensor current as a function of two adjacent plunger gate voltages. 
Charge state transition lines can be observed in a CSD, separating the diagram into four main regions, each representing a distinct charge state. The charge state of a dot is simply the number of electrons trapped in a dot. In the case of a double dot, we use a tuple $(N_1, N_2)$ to represent the double dot's charge state collectively, where $N_1$ is the number of electrons in dot 1, and $N_2$ is the number of electrons in dot 2.
The charge states of the four regions are labeled in 
Figure~\ref{fig:dqd-charge-stability}.
Assuming we start from the $(0, 0)$ region, as $P_2$ voltage is increased, the charge state changes from $(0, 0)$ to $(0, 1)$, corresponding to the arrow \circled{2} in Figure~\ref{fig:dqd-charge-stability} (right), meaning that the increased $P_2$ voltage traps an electron in dot 2. Similarly, increasing the $P_1$ voltage achieves transition from $(0, 0)$ to $(1, 0)$, corresponding to the arrow \circled{1} in Figure~\ref{fig:dqd-charge-stability}, traps an electron in dot 1.

The cross-capacitance effect is indicated by the slopes of the charge state transition lines in the CSD. 
Assuming we start from a point in the $(0, 1)$ region and close to the $(1, 1)$ region, then due to the sloped $(0, 1)$ to $(1, 1)$ transition line, 
an increment in $P_2$ voltage may transit the state to $(1, 1)$, represented by the short arrow \circled{3} in Figure~\ref{fig:dqd-charge-stability}. 
This is an undesirable effect, and ideally, plunger gate voltages should independently control the charge state of its corresponding dot.

\subsection{Virtual Gate}
To resolve the aforementioned cross-capacitance nonideality and realize 
precise 
orthogonal control over individual quantum dots, a virtual gate technique is introduced to 
compensate for the non-negligible side effects from nearby plunger gates.
A dot's virtual gate voltage is a linear combination of nearby physical gate voltages with carefully determined coefficients such that the linear combination of them realizes the ideal "one-to-one" control of the dot's potential, where one can adjust the energy level of the target dot without affecting adjacent dots.

\begin{figure}[t]
  \centering
  \includegraphics[width=0.9\columnwidth]{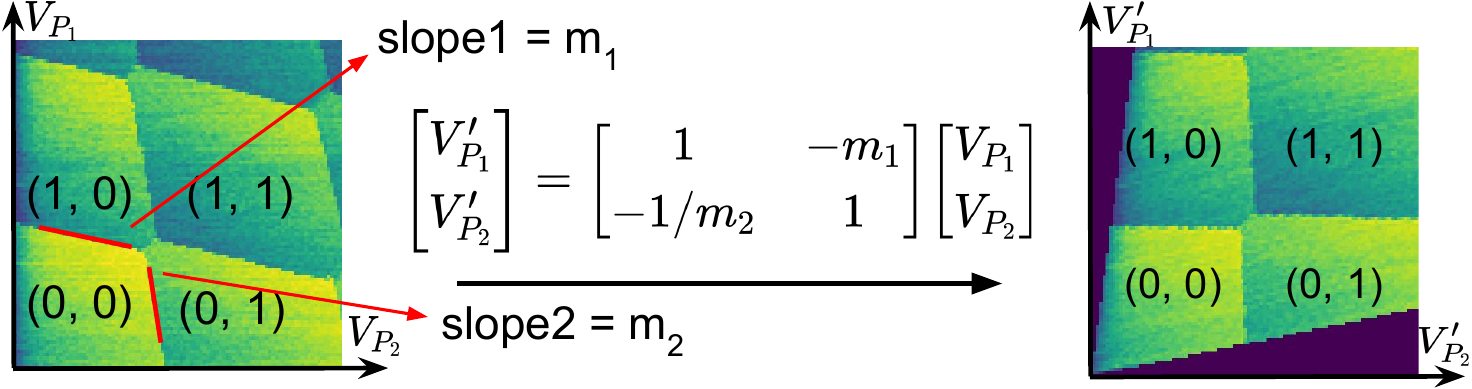}
  \vspace{-8pt}
  \caption{Virtual gate example}
  \label{fig:vg-effect}
\end{figure}

Figure~\ref{fig:vg-effect} shows an example of virtual gates for a double quantum dot device.
On the left is the original CSD. $V_{P_1}$ and $V_{P_2}$ are the actual voltages on plunger gates 1 and 2.
The virtual gate voltages $V_{P_1}^\prime$ and $V_{P_2}^\prime$ are the linear combination of the actual gate voltages:
\vspace{-5pt}
\[\small
\setlength{\tabcolsep}{10pt} 
\renewcommand{\arraystretch}{0.9} 
\begin{bmatrix}
  V_{P_1}' \\
  V_{P_2}'
\end{bmatrix}
=
\begin{bmatrix}
  1 & \alpha_{12} \\
  \alpha_{21} & 1
\end{bmatrix}
\begin{bmatrix}
  V_{P_1} \\
  V_{P_2}
\end{bmatrix}
\]
where the 2x2 matrix is the virtualization matrix. The process of establishing virtual gates is equivalent to finding $\alpha_{12}$ and $\alpha_{21}$ in the virtualization matrix.

The values of $\alpha_{12}$ and $\alpha_{21}$ can be computed based on the slopes of the transition lines - the boundaries between different charge state areas in a CSD.
Suppose $m_1$ and $m_2$ are the slopes of (0, 0)\textrightarrow (0, 1) and (0, 0)\textrightarrow (1, 0) transition lines.
Then we have \vspace{-3pt}
$$\small \setlength{\tabcolsep}{10pt} 
\renewcommand{\arraystretch}{1.5} 
\alpha_{12} = -m_1, \quad \alpha_{21} = -1/m_2$$ 
The virtual gate extraction can be extended to a $n$-dot array by sequentially applying it to every pair of nearby plunger gates~~\cite{Mills3}, and $n-1$ sequentially executed extraction processes are needed for an n-dot array.

On the right of Figure~\ref{fig:vg-effect} is the CSD with the virtual gates. The virtualization matrix defines an affine transformation that warps the space to orthogonalize the transition lines. The orthogonal transition lines in the virtualized space demonstrate "one-to-one" control---increasing one 
virtual plunger gate voltage only affects the charge state on its corresponding dot.

\section{Previous Work}

Previous works for constructing the virtualization matrix are mostly obtaining the full CSD for each pair of nearby plunger gates and then applying computer vision techniques on the CSDs to extract the slopes of the transition lines~\cite{Mills2, Giovanni, Mills3, Ziegler}.
However, obtaining a full CSD is slow. Each data point $(V_{P_1}, V_{P_2})$ in the CSD requires physically setting the voltages of the two plunger gates to be $V_{P_1}$ and $V_{P_2}$, waiting a dwell time,  and then measuring the current in the charge sensor (shown as a function in Algorithm~\ref{alg:getcurrent}). 

\begin{algorithm}[t]\small
\setlength{\tabcolsep}{10pt} 
\renewcommand{\arraystretch}{0.9} 
\caption{Get charge sensor current}\label{alg:getcurrent}
\begin{algorithmic}[1]
\Function{getCurrent}{$v_1$, $v_2$}
\State Set gate voltages to $v_1$, $v_2$
\State $wait(\text{dwellTime})$
\State \Return  charge sensor current 
\EndFunction
\end{algorithmic}
\end{algorithm}
\setlength{\textfloatsep}{2pt}

The major bottle neck is the dwell time in line 3. This delay is necessary to account for the heavy filtering on the wires applying the bias voltages to the device. While hardware improvements can reduce this dwell time, typical dwell times are often on the order of milliseconds, meaning typically it takes minutes to record one CSD of two quantum dots. Due to this overhead, the CSD acquisition becomes significant overhead for tuning large-scale devices. 

\section{Fast Virtual Gate Extraction}

This paper aims to accelerate the virtual gate extraction so that the electron spin qubits can be quickly established for further experiments.
Our optimization is based on a key observation that there is significant redundancy in the existing computer vision-based virtual gate extraction process.
We aim to locate the transition lines using as few data points in the CSD as possible.
Then, in the actual experiments, we only need to probe very few data points to reconstruct the virtualization matrix.
\vspace{-5pt}

\subsection{Redundancy in Full CSD Data}

Since constructing the virtualization matrix depends on the slopes of the transition lines,
only the data points near the transition lines will contribute to the calculation.
Ideally, to find the slope of a line, if there were no noise in the experiment, two data points along the line would be enough to determine the slope of one transition line.
Although there is inevitable noise in real experiments and more data points are required to precisely determine the transition lines,
a lot of data points in the CSD are not necessary 
since the transition lines and their nearby data points are only a small portion of the entire CSD. 
\vspace{-5pt}

\subsection{Locating Critical Region}

\begin{figure}[b]
    \centering
    \vspace{5pt}
    \includegraphics[width=0.8\columnwidth]{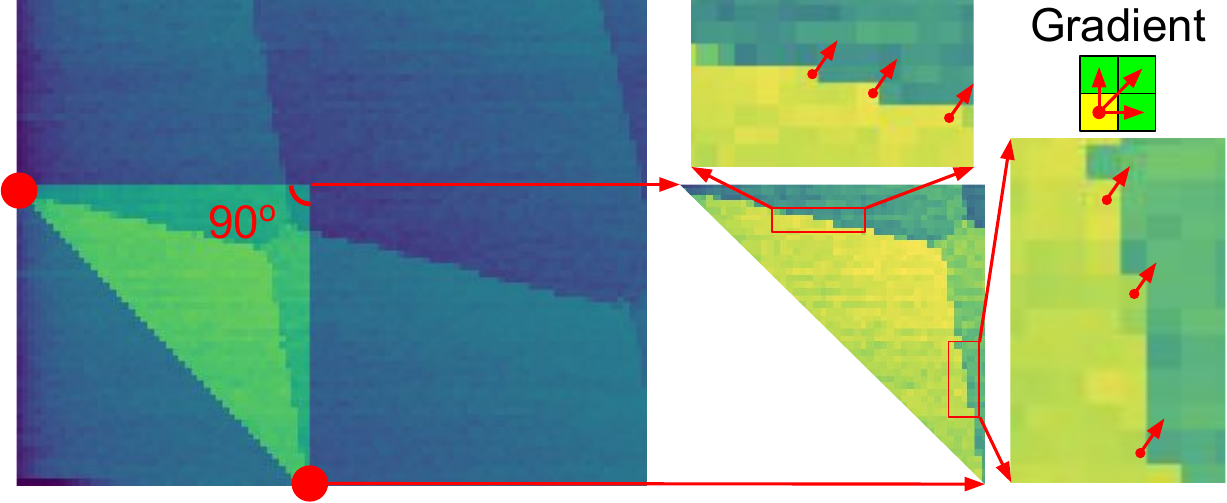}
    \vspace{-7pt}
    \caption{The highlighted area confines both transition lines. Points on the transition lines have a larger gradient in the positively tilted direction}
    \label{fig:fig4}
\end{figure}

To eliminate the redundancy, the first step of our method is to locate the region where the transition lines will lie in roughly.
We first observe that the transition lines have negative slopes, and the slope of the \mbox{(0, 0)\textrightarrow (0, 1)} transition line is steeper than the \mbox{(0, 0)\textrightarrow (1, 0)} transition line.
This observation can be validated by the physics capacitance model for quantum dots \cite{Hanson}. 
This prior knowledge suggests that if we know two endpoints (one on the \mbox{(0, 0)\textrightarrow (1, 0)} transition line and one on the \mbox{(0, 0)\textrightarrow (0, 1)} transition line), both transition lines are within the right triangular area defined by these two points. 
An example is shown in Figure~\ref{fig:fig4}.
If the two red points are known,  the highlighted right triangular region can cover both transition lines, and we only need to search in this region to locate the rest of the points on the lines. We name the points defining the triangle ``anchor points''.
We assume that the anchor points are known for now and will introduce how to find them later in Section~\ref{sec:anchor}.

\subsection{Searching in the Critical Region}
\subsubsection{Gradient-Based Feature}
After locating the critical region, we use a gradient-based feature to locate the points on the transition lines inside the triangular area. 
Note that a transition line indicates a sharp change in charge sensor current across the line.
This feature results in a positively tilted gradient at the transition points, as shown on the top right of Figure~\ref{fig:fig4}. 
We use this positively tilted gradient as the feature to find the transition lines.
For a point in the voltage space, we represent its positively tilted gradient by the sum of its differences in charge sensor current with the right and upper-right pixels, which we define as the feature gradient. Note the pixel here refers to the voltage granularity, not the actual image pixel. On real devices, assuming the gate voltages are at ($V_{P_1}$, $V_{P_2}$), to get the charge sensor current at the right and upper-right pixels, the gate voltages are sequentially adjusted to  
($V_{P_1}$, $V_{P_2} + \delta$)
and ($V_{P_1} + \delta$, $V_{P_2} + \delta$), where $\delta$ is the voltage granularity that defines the pixel size. The procedure for computing gradient at gate voltages ($v_1$, $v_2$) is shown in Algorithm~\ref{alg:gradient}.

\begin{algorithm}[b]\small
\setlength{\tabcolsep}{10pt} 
\renewcommand{\arraystretch}{0.9} 
\caption{Compute feature gradient}\label{alg:gradient}
\begin{algorithmic}[1]
\Function{GetGradient}{$v_1$, $v_2$}
\State $c \gets \textsc{GetCurrent}(v_1, v_2)$
\State $cRight \gets \textsc{GetCurrent}(v_1, v_2 + \delta)$
\State $cUpperRight \gets \textsc{GetCurrent}(v_1 + \delta, v_2 + \delta)$
\State \Return 
$(c - cRight) + (c - cUpperRight)$
\EndFunction
\end{algorithmic}
\end{algorithm}
\vspace{-10pt}

\subsubsection{Search for Transition Lines}
Using the gradient-based feature above, our method will locate points on the charge state transition lines in the triangular region by performing a bottom-to-top row-major sweep and a left-to-right column-major sweep. During the sweeps, the triangular region is dynamically shrunk to ensure only points near the transition lines are probed.

\begin{figure}[t]
  \centering
  \begin{subfigure}{0.49\columnwidth}
    \centering
    \includegraphics[width=0.85\linewidth]{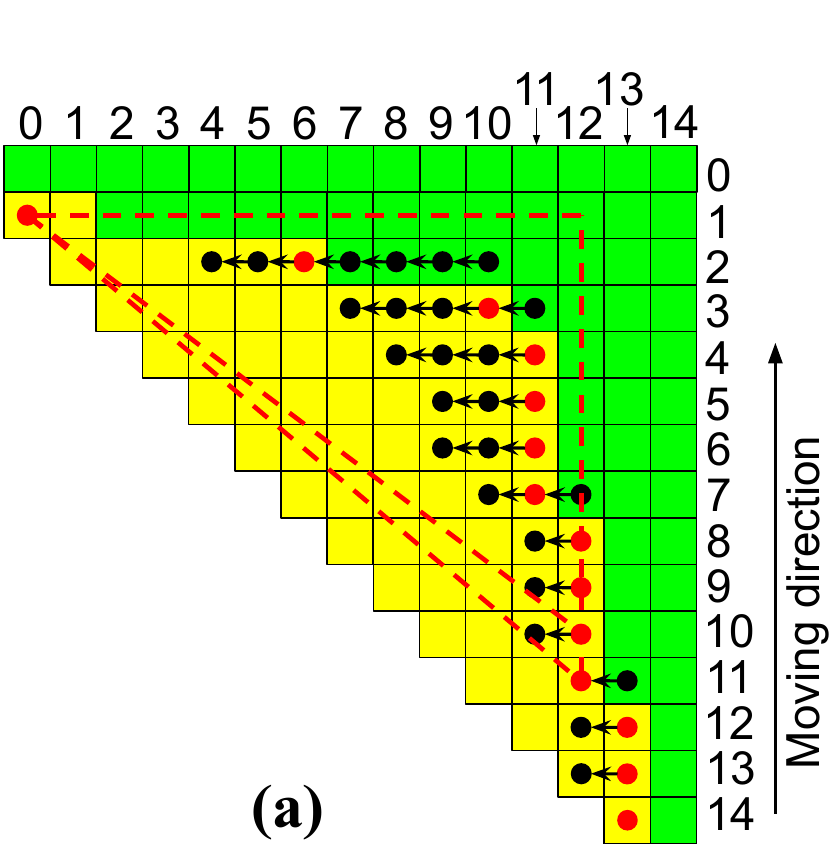}
  \end{subfigure}
  \hfill
  \begin{subfigure}{0.49\columnwidth}
    \centering
    \includegraphics[width=0.85\linewidth]{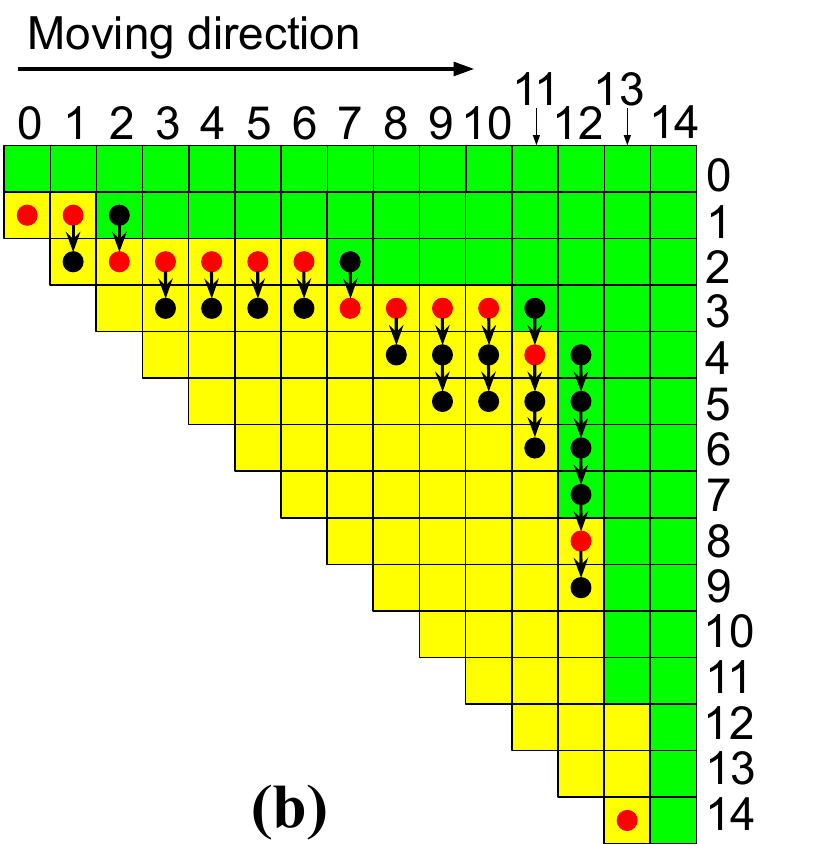}
  \end{subfigure}
    \vspace{-5pt}
  \caption{(a) Row-major sweep, (b) Column-major sweep}
  \label{fig:sweeps}
\end{figure}

The row-major sweep goes through the points in a row-major fashion. As shown in Figure~\ref{fig:sweeps} (a), the sweep iterates through the rows from bottom to top. The upper left anchor point is fixed, and the lower right anchor point moves up. At each row, it probes the segment of points within the triangular region. We use a point's pixel center to determine whether it's inside the triangular region. \textsc{GetGradient} is applied to each point in the segment to calculate their feature gradient, and the point with the maximum feature gradient is saved as a charge state transition point. We then update the lower anchor point to be the newly found charge state transition point--effectively shrinking the triangular area. Figure~\ref{fig:sweeps} (a) demonstrates the row-major sweep. The grid represents the voltage space. The points probed are marked with spots. The red spots represent the anchor points on each row. 
The black spots represent the points probed but not saved. 
The small black arrows on each row represent the order and direction of the sweep.
As an example, when sweeping row 10, the following procedures are applied: 
\setlength{\itemsep}{1pt}
\begin{enumerate}[label=\arabic*.]
  \addtocounter{enumi}{0}
  \vspace{-4pt}
  \item The lower anchor point is currently at (11, 12) (row, column), which is the charge state transition point located at row 11. The other fixed anchor point is at (1, 0). These two anchor points define the right triangular region whose vertices are (1, 0), (1, 12), (11, 12). The edges of the triangle are the dashed red lines in Figure~\ref{fig:sweeps} (a).
  \item The sweep goes through points (10, 12) and (10, 11), the only two points within the triangular area on row 10, and then computes their feature gradients. Here, point (10, 12) has a larger feature gradient, so it is saved as a charge state transition point.
  \item Update the lower anchor point from (11, 12) to the newly found charge state transition point (10, 12). \vspace{-4pt}
\end{enumerate}
The row-major sweep performs these procedures on each row until it reaches the row of the fixed anchor point (row 1 in 5a). Notice that the shrinking triangular region keeps the search near the transition lines, probing only a few points at each row where charge state transition is most likely to occur. 

The row-major sweep is effective in locating points on the (0, 0)\textrightarrow (0, 1) transition line but less effective in locating points on the \mbox{(0, 0)\textrightarrow (1, 0)} transition line because it is less orthogonal to the sweeping direction. As shown in Figure~\ref{fig:sweeps} (a), when the row-major sweep reaches the \mbox{(0, 0)\textrightarrow (1, 0)} transition region (row 3 and above), the segment inside the triangular region becomes relatively long (e.g., the segment at row 2 in Figure~\ref{fig:sweeps} (a) is longer than the segments at row 5-13), making the search more susceptible to noise. It is also possible in this region that a falsely located point deviates the triangular region from the transition line, producing a series of falsely located points. To locate equally accurate points on the \mbox{(0, 0)\textrightarrow (1, 0)} transition line, a column-major sweep is performed in analogs to the row-major sweep. The column-major sweep is performed from left to right and is demonstrated in Figure~\ref{fig:sweeps} (b). It is similar to the previous row-major sweep, with only the row and column exchanged.

\begin{figure}[b]
  \centering
  \vspace{3pt}
  \includegraphics[width=0.9\columnwidth]{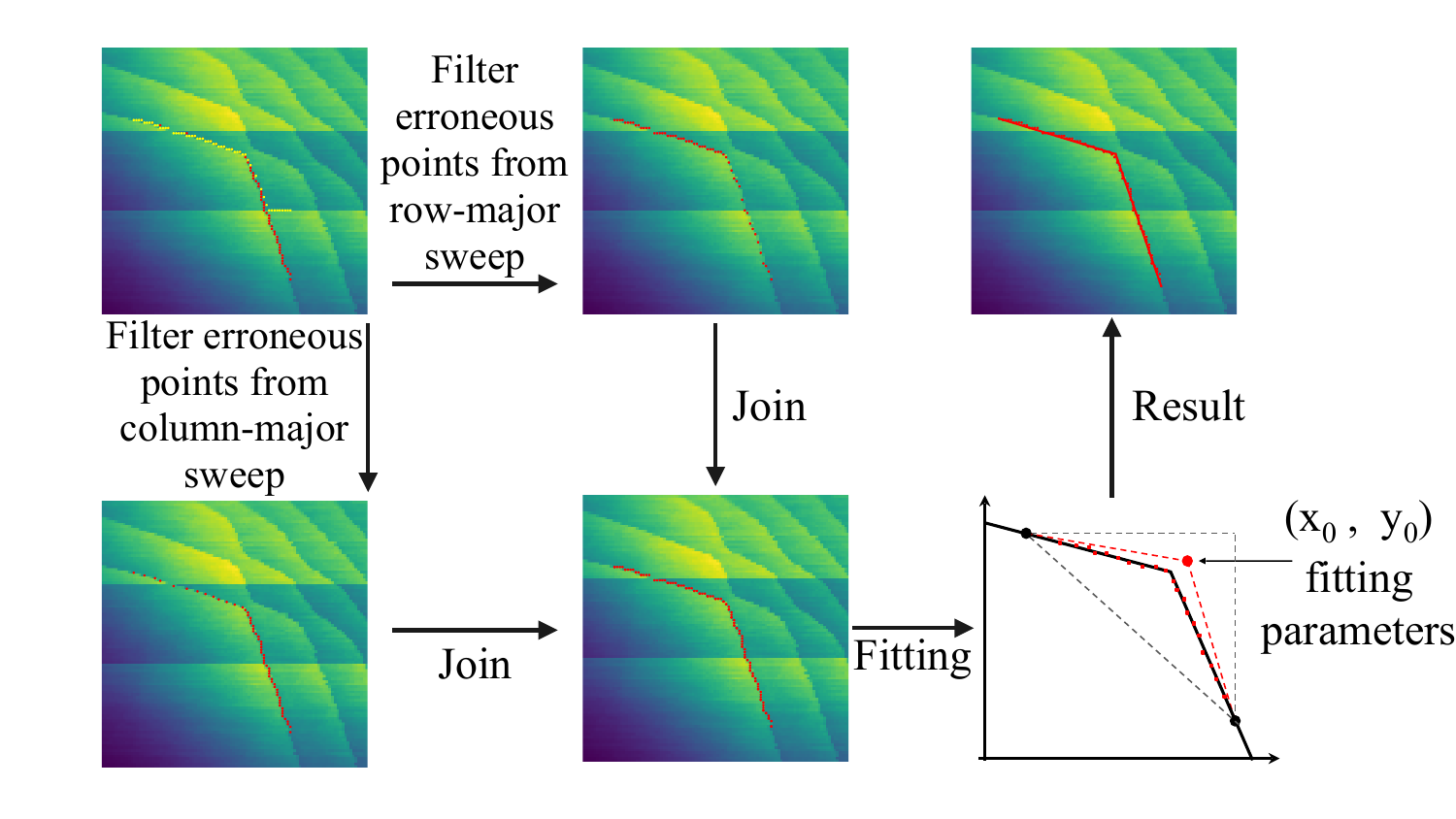}
  \vspace{-9pt}
  \caption{Post processing procedure running on example data}
  \label{fig:postpreprocessing}
\end{figure}

A post-processing step shown in Figure~\ref{fig:postpreprocessing} is included to filter erroneous points.  
The upper left image in Figure~\ref{fig:postpreprocessing}  shows all points located by the two sweeps on an example voltage space, where the red points are produced by the row-major sweep and the yellow points are produced by the column-major sweep. 
Erroneous points are likely to be produced when the row-major sweep attempts to locate points in the \mbox{(0, 0)\textrightarrow (1, 0)} transition region and when the column-major sweep attempts to locate the points in the \mbox{(0, 0)\textrightarrow (0, 1)} transition region because they will sweep more points in a row/column and are more error-prone. The post-processing step produces two filtered sets of points and joins them together. For the first set, we only keep the point with the smallest y value at each x position (i.e., the lowest point in each column). This is designed to exclude erroneous red points using the yellow points below them. Similarly, for the second set, we only keep the point with the smallest x value at each y position (i.e., the leftmost point in each row). 
By joining the two filtered sets, we get a comprehensive set of points on both transition lines with minimal erroneous points. The procedures of row-major, column-major sweeps, and post-processing are summarized in Algorithm~\ref{euclid}.

\setlength{\floatsep}{15pt}
\begin{algorithm}[t]\small
\setlength{\tabcolsep}{10pt} 
\renewcommand{\arraystretch}{0.9} 
\caption{Sweeps + Post-processing}\label{euclid}
\begin{algorithmic}[1]
\Procedure{PostProcess}{$points$}
\State $filteredPoints1 = 
\{(x, y) \mid \forall (x, y') \in points, y \leq y' \}$
\State $filteredPoints2 = 
\{(x, y) \mid \forall (x', y) \in points, x \leq x' \}$
\State \Return $filteredPoints1 \cup filteredPoints2$
\EndProcedure
\Procedure{sweeps}{$(x_1, y_1), (x_2, y_2)$}
\State $anchorPoint1, anchorPoint2  \gets (x_1, y_1), (x_2, y_2)$
\State $transitionPoints = \left[\right]$
\For {$i \gets y_2 + 1$ to $y_1 - 1$ }
  \State $P \gets \text{points inside the triangular region at row } i$
  \State $(x^*, y^*) \gets 
  \underset{(x, y) \in P}{\operatorname{argmax}} \;
  \textsc{getGradient}(y, x)$
  \State append $(x^*, y^*)$ to $transitionPoints$
  \State $anchorPoint2 \gets (x^*, y^*)$
\EndFor
\State $anchorPoint2 \gets (x_2, y_2)$
\For {$j \gets x_1 + 1$ to $x_2 - 1$}
  \State $P \gets \text{points inside the triangular region at column } j$
  \State $(x^*, y^*) \gets 
  \underset{(x, y) \in P}{\operatorname{argmax}} \;
  \textsc{getGradient}(y, x)$
  \State append $(x^*, y^*)$ to $transitionPoints$
  \State $anchorPoint1 \gets (x^*, y^*)$
\EndFor
\State \Return $\textsc{postProcess}(transitionPoints)$
\EndProcedure
\end{algorithmic}
\end{algorithm}
\setlength{\textfloatsep}{2pt}

\subsubsection{Slope Extraction}
After the transition line points are located, we parameterize a 2-piece-wise linear shape by the two initial anchor points and the intersecting point of the two transition lines, where the coordinates of the intersecting point are the fitting parameters. We use SciPy's curve fitting function to find the optimal position of the intersecting point~\cite{2020SciPy-NMeth}. The slopes of the transition lines are then computed using the intersecting point and the initial anchor points.

\subsection{Preprocessing for Anchor Points}
\label{sec:anchor}

Finally, we introduce our preprocessing method to locate the two initial anchor points. 
We first probe ten equally spaced points spanning the diagonal from the lower left to the upper right.
Then we apply the following masks along x and y directions, starting from the brightest point found in the previous step or 10\% width and height, whichever is more distant from the lowest and leftmost point. These masks are designed to compute a positively sloped gradient across three pixels, a more noise-resilient indicator of charge state transition than the feature gradient. 
\setlength{\abovedisplayskip}{2pt}
\setlength{\belowdisplayskip}{2pt}
$$
Mask_x=
\begin{bmatrix}
    1 & 1 & -3 & -4 & -4\\
    2 & 2 &  0 & -2 & -2\\
    4 & 4 &  3 & -1 & -1
\end{bmatrix}, \
Mask_y = 
\begin{bmatrix}
    -1 & -2 & -4 \\
    -1 & -2 & -4 \\
    3  &  0 & -3 \\
    4 & 2 & 1 \\
    4 & 2 & 1 \\
\end{bmatrix}
$$
We sweep the masks along their corresponding axis and sum the entries of their element-wise product with the pixel values. The resulting arrays are then multiplied element-wise by the 1D Gaussian distribution. The point with the maximum value in each array is used as an initial anchor point.


\section{Evaluation}

In this section, we evaluate the proposed fast virtual gate extraction method with respect to the speed and accuracy of experimental data of real quantum dot devices.

\begin{table*}[t]
\setlength{\tabcolsep}{10pt} 
\renewcommand{\arraystretch}{1} 
\caption{Result Summary}
\vspace{-10pt}
\resizebox{0.9\textwidth}{!}{ 
\begin{tabular}{cc | cc | cc | cc | c}
\multicolumn{2}{c|}{Benchmark} &
\multicolumn{2}{c|}{Success/Fail} &
\multicolumn{2}{c|}{Number/percentage of points probed} &
\multicolumn{2}{c|}{Total runtime} &
Speedup\\
    \hline
CSD Index& Size & Fast Extraction & Baseline & Fast Extraction & Baseline & Fast Extraction & Baseline & Speedup\\

1& 200$\times$200 & Fail & Fail & 1753 (4.38\%) & 40000 (100\%)& 87.94s& 2005.12s& N/A\\
2& 200$\times$200 & Fail & Fail & 1692 (4.23\%)& 40000 (100\%)& 84.89s& 2004.98s & N/A\\
3& 63$\times$63 & Success & Success & 643 (16.2\%)& 3969 (100\%)& 32.26s& 198.96s& 6.16$\times$\\
4& 63$\times$63 & Success & Success & 679 (17.1\%)& 3969 (100\%)& 34.06s& 198.97s& 5.84$\times$\\
5& 63$\times$63 & Success & Success & 484 (12.19\%)& 3969 (100\%)& 24.28s& 198.94s& 8.19$\times$\\
6& 100$\times$100 & Success & Success & 1002 (10.02\%)& 10000 (100\%)& 50.27s& 501.25s& 9.97$\times$\\
7& 100$\times$100 & Success & Fail & 985 (9.85\%)& 10000 (100\%)& 49.42s& 501.26s& 10.14$\times$\\
8& 100$\times$100 & Success & Success & 1179 (11.79\%)& 10000 (100\%)& 59.14s& 501.26s& 8.48$\times$\\
9& 100$\times$100 & Success & Success &974 (9.74\%)& 10000 (100\%)& 48.86s& 501.27s & 10.26$\times$\\
10& 100$\times$100 & Success & Success & 1054 (10.54\%)& 10000 (100\%)& 52.88s& 501.27s& 9.48$\times$\\
11& 100$\times$100 & Success & Success & 927 (9.27\%)& 10000 (100\%)& 46.5s& 501.26s & 10.78$\times$\\
12& 200$\times$200 & Success & Success & 2067 (5.17\%)& 40000 (100\%)& 103.69s& 2005.02s& 19.34$\times$\\
\end{tabular}}
\vspace{-15pt}
\label{tab:result}

\end{table*}

\subsection{Experiment Setup}
\textbf{Benchmark}
We adopt the 12 experimentally measured charge stability diagrams (CSDs) in the qflow dataset version 2~\cite{qflow}. Note that this has covered all the real experimental data from this dataset. 
These CSDs were measured in the double-dot configuration on a triple-dot $Si/SiGe$ device fabricated on a 300mm industrial line~\cite{qflow, testDataDevice}. The diagrams are cropped to focus on the 50\% width and height region where the (0, 0), (0, 1), (1, 0), (1, 1) charge state regions are located. The final data have pixel resolutions ranging from 63 $\times$ 63 (126$\times$126 before cropping) to 200$\times$200 (401$\times$401 before cropping).

\textbf{Metrics} We use the rate of successfully finding the transition lines to indicate the applicability of the proposed algorithm and the baseline.
We use the total runtime to evaluate the overall speedup.
We also collect the number of data points probed because this is the bottleneck in the entire virtual gate extraction process, and our speed is achieved by reducing the number of probed data points.

\textbf{Implementation}
We prototype the proposed virtual gate extraction algorithm in Python and simulate 
the experiment setup using the data from real experiments. 
When the proposed algorithm needs to obtain a data point with a specific voltage combination, it will call a simulated \hbox{\textsc{getCurrent}} function with the voltage coordinates.
The \hbox{\textsc{getCurrent}} function will return a current from a CSD in the dataset to simulate an experiment on that device.
The \hbox{\textsc{getCurrent}} function uses a dwelling time of 50ms, which is a typical time delay for charge-sensor-based devices~\cite{ZajacThesis}.
Since the benchmark dataset does not provide reference virtual gates, we plot the final affine transformed diagram for virtualization and manually examine 
whether the virtual gate extraction is successful or not.

\textbf{Baseline}
Our baseline is the Hough transform with Canny edge detection, an existing technique used for virtual gate automation~\cite{Mills2, Giovanni} implemented using the opencv-python~\cite{opencv_library}. The baseline first obtains a full CSD by calling the simulated \textsc{getCurrent} on all voltage combinations before image processing. 


\subsection{Result}

Table~\ref{tab:result} shows the results of applying the proposed method and the baseline on all the benchmarks.
Overall, our fast virtual gate extraction method can successfully find the transition lines and then construct the virtual gate in 10 out of the 12 benchmarks, while the baseline Hough transform succeeds on only 9 benchmarks.
We manually checked the two benchmarks we failed on.
The reason is that those two devices have too much noise in the CSDs and both our method and the baseline fail to locate the transition lines.
We also investigated the case CSD 7 where the baseline fails while our method succeeds.
We found that the edge detection in the baseline could not locate enough points to establish the line. 
Meanwhile, the column-major sweep in our method successfully located the line.

\begin{figure}[h]
  \hspace{0.05\columnwidth}\begin{subfigure}{0.3\columnwidth}
    \centering
    \includegraphics[width=0.85\linewidth]{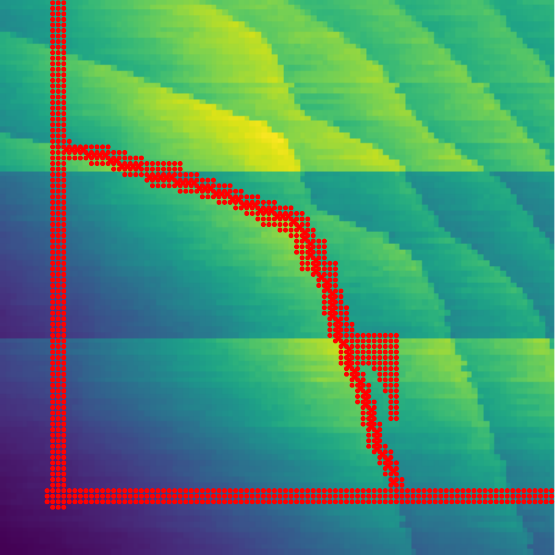}
  \end{subfigure}
  \hspace{0.15\columnwidth}
  \begin{subfigure}{0.3\columnwidth}
    \centering
    \includegraphics[width=0.85\linewidth]{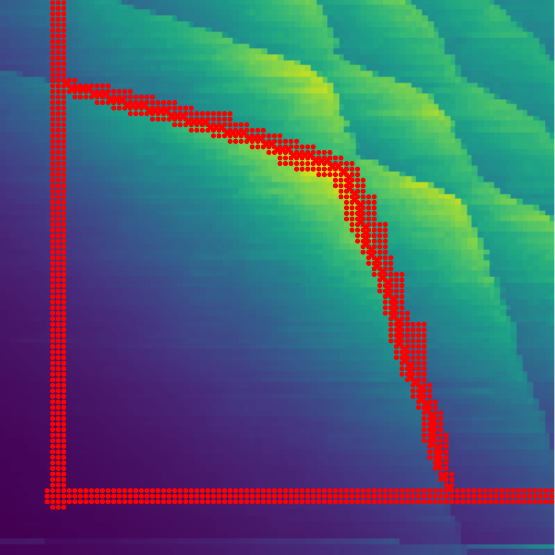}
  \end{subfigure}
  \vspace{-6pt}
  \caption{Data points probed in benchmark CSD 6 and 10}
  \vspace{-10pt}
  \label{fig:datapointprobe}
\end{figure}

In terms of the virtual gate extraction speed, our fast extraction method can achieve a speedup ranging from $5.84\times$ to $19.34\times$ against the baseline.
The speed-up mostly comes from the reduction in the number of data points probed as our method adaptively locates the critical region and only probes a small number of actual gate voltage configurations.
The baseline probes every point in the space to fill a full CSD, while our method only probes $10\%$ on average as shown in the middle of Table~\ref{tab:result}.
As an example, Figure~\ref{fig:datapointprobe} shows the data points probed in CSD 6 and CSD 10 by our method, and the probed data points for other benchmarks are similar. Note that the points are mostly scattered around the two transition lines, with additional points probed to determine the initial anchor points. 
In summary, our fast virtual gate extraction method outperforms the Hough transform baseline with faster virtual gate extraction, fewer experimental data requirements, and even higher accuracy.

\vspace{-6pt}
\setlength{\textfloatsep}{2pt}

\vspace{-5pt}
\begin{acks}
    This work was supported in part by NSF CAREER Award 2338773.
\end{acks}

\vspace{-5pt}
\bibliographystyle{ACM-Reference-Format}
\bibliography{base}

\end{document}